\documentclass[manuscript=article]{achemso}

\usepackage{graphicx}
\usepackage{color}
\usepackage{mathrsfs}
\usepackage{bm}
\usepackage[version=3]{mhchem}
\usepackage{color, soul}
\usepackage[english]{babel}
\addto\captionsenglish{}
\addto\captionsenglish{}
\graphicspath{{figures/}}
\usepackage{siunitx}
\usepackage[utf8]{inputenc}
\usepackage{miller}
\usepackage{lineno}

\title{Continuous-Wave Second-Harmonic Generation in Orientation-Patterned {\color{black}Gallium Phosphide} Waveguides at Telecom Wavelengths}
\author{Konstantinos Pantzas}
\affiliation{Centre de Nanosciences et de Nanotechnologies, CNRS, Université Paris Saclay,91120 Palaiseau, France}
\email{konstantinos.pantzas@c2n.upsaclay.fr}
\author{Sylvain Combrié}
\affiliation{Thales Research and Technology, Campus Polytechnique, 1 avenue Augustin Fresnel, 91767 Palaiseau, France}
\author{Myriam Bailly}
\affiliation{Centre de Nanosciences et de Nanotechnologies, CNRS, Université Paris Saclay,91120 Palaiseau, France}
\altaffiliation{Thales Research and Technology, Campus Polytechnique, 1 avenue Augustin Fresnel, 91767 Palaiseau, France}
\author{Raphaël Mandouze}
\affiliation{Thales Research and Technology, Campus Polytechnique, 1 avenue Augustin Fresnel, 91767 Palaiseau, France}
\author{Francesco Rinaldo Talenti}
\affiliation{Thales Research and Technology, Campus Polytechnique, 1 avenue Augustin Fresnel, 91767 Palaiseau, France}
\altaffiliation{Dipartimento di Ingegneria dell’Informazione, Elettronica e Telecomunicazioni, Sapienza University of Rome, 00184 Rome, Italy}
\author{Abdelmounaim Harouri}
\affiliation{Centre de Nanosciences et de Nanotechnologies, CNRS, Université Paris Saclay,91120 Palaiseau, France}
\author{Bruno Gérard}
\affiliation{III-V Lab, Campus Polytechnique, 1 avenue Augustin Fresnel, 91767 Palaiseau, France}
\author{Grégoire Beaudoin}
\affiliation{Centre de Nanosciences et de Nanotechnologies, CNRS, Université Paris Saclay,91120 Palaiseau, France}
\author{Luc Le Gratiet}
\affiliation{Centre de Nanosciences et de Nanotechnologies, CNRS, Université Paris Saclay,91120 Palaiseau, France}
\author{Gilles Patriarche}
\affiliation{Centre de Nanosciences et de Nanotechnologies, CNRS, Université Paris Saclay,91120 Palaiseau, France}
\author{Alfredo De Rossi}
\affiliation{Thales Research and Technology, Campus Polytechnique, 1 avenue Augustin Fresnel, 91767 Palaiseau, France}
\author{Yoan Léger}
\affiliation{Institut Foton (UMR6082/CNRS-Univ. Rennes-INSA), 20 avenue des Buttes de Coësmes, 35708 Rennes, France}
\author{Isabelle Sagnes}
\affiliation{Centre de Nanosciences et de Nanotechnologies, CNRS, Université Paris Saclay,91120 Palaiseau, France}
\author{Arnaud Grisard}
\affiliation{Thales Research and Technology, Campus Polytechnique, 1 avenue Augustin Fresnel, 91767 Palaiseau, France}

\begin{document}

\begin{abstract}
A new process to produce Orientation-Patterned Gallium Phosphide (OP-GaP) on GaAs with almost perfectly parallel domain boundaries is presented. Taking advantage of the chemical selectivity between phosphides and arsenides, OP-GaP is processed into suspended shallow-ridge waveguides. Efficient Second-Harmonic Generation from Telecom wavelengths is achieved in both Type-I and Type-II polarisation configurations. The highest observed conversion efficiency is \SI{200}{\percent\per\watt\per\centi\meter\squared}, with a bandwidth of \SI{2.67}{\nano\meter} in a \SI{1}{\milli\meter}-long waveguide. The variation of the conversion efficiency with wavelength closely follows a squared cardinal sine function, in excellent agreement with theory, confirming the good uniformity of the poling period over the entire length of the waveguide.
\end{abstract}


\section{Introduction}

Global telecommunication networks are built on the optical-fiber transparency windows at \SI{1.3}{\micro\meter} and \SI{1.55}{\micro\meter}, where transmission losses are minimal, driving the development of state of the art lasers, amplifiers and modulators at these same wavelengths. Best in class detectors, on the other hand, have been developed at visible-light wavelengths, where Si CMOS detectors offer both excellent performance and unmatched cost-effectiveness. An ideal telecommunications chain would combine emission and transmission at telecom wavelengths with detection at visible wavelengths. Nonlinear optical processes, for instance frequency upconversion through second harmonic generation, are a promising solution to bridge this gap, at the condition, nevertheless, that the nonlinear process is efficient.

Achieving high nonlinear conversion efficiencies is no mean feat. Though a large second-order nonlinear coefficient is a prerequisite, phase-matching is equally important - if not more: linear dispersion in solids varies significantly with wavelength, inducing a phase shift between the fundamental and the converted mode. This phase shift periodically extinguishes the nonlinear conversion process, making ideal phase matching unattainable in practice. Quasi phase matching (QPM) in periodically-poled nonlinear media is a proven alternative to overcome this hurdle. Indeed, optical parametric oscillators (OPOs) based on MgO-doped periodically-polled \ce{LiNbO3} (PPLN) are firmly established as high-power and widely tunable sources between \SI{1}{\micro\meter} and \SI{4}{\micro\meter}. Record upconversion efficiencies, exceeding \SI{2500}{\percent\per\watt\per\centi\meter\squared}, have also been demonstrated using PPLN waveguides \cite{wang2018ultrahigh}.  However, PPLN suffers from drawbacks such as {\color{black}strong Raman effect \cite{Yu:2020uz}} and a strong photorefractive effect \cite{Luedtke:09,Mondain:20}. Furthermore, while significant advances have been made, these oxide crystals are not as readily integrated in Si-photonics as Group IV or III-V semiconductors \cite{Liu:15, Takenaka:2017jstqe,Saleem-Urothodi:20,Roland:2016um}.

Addressing these issues has prompted the development of non-oxide-based nonlinear crystals such as orientation-patterned gallium arsenide (OP-GaAs). In orientation-patterned crystals, the $\chi^{(2)}$ is periodically reversed by reversing the polarity of the semiconductor crystal. OP-GaAs has been very successful replacing PPLN at wavelengths between \SI{4}{\micro\meter} and \SI{12}{\micro\meter}. Nevertheless, the short-wavelength absorption-edge in these crystals typically lies around \SI{2}{\micro\meter}. Two-photon absorption is particularly strong at telecom wavelengths in GaAs and precludes OP-GaAs crystals from being used as the nonlinear medium for second harmonic generation at \SI{1.55}{\micro\meter}.

This is where gallium phosphide (GaP) comes in. Indeed, GaP has negligible two-photon absorption for wavelengths above \SI{1}{\micro\meter}, in addition to several other properties that make it alluring for nonlinear photonics: large $\chi^{(2)}$ and $\chi^{(3)}$ nonlinearities, a large susceptibility (\SI{50}{\pico\meter\per\volt}), and high transparency from \SI{500}{\nano\meter} to \SI{12}{\micro\meter} \cite{levine1972nonlinear,levine1991calculation,shoji2002second}. Moreover, GaP has a small lattice mismatch with Si (\SI{0.1}{\percent} at room temperature), and exhibits a high thermal conductivity (\SI{110}{\watt\per\meter\per\kelvin}), making it an excellent candidate for integrated nonlinear photonics. As a result, several groups have recently made efforts to harness these properties and develop OP-GaP crystals.

Three different approaches have been used to obtain the OP-GaP crystals. All approaches rely on the fabrication of a seed layer, wherein the orientation-reversal pattern is defined. The first consists in using an OP-GaAs seed  \cite{schunemann2021continuous,wei2018performance,tassev2016heteroepitaxial}, the fabrication of which is technologically mature. In the other two approaches the seed is fabricated using bulk GaP. Orientation reversal is obtained either through wafer bonding  \cite{TASSEV201272}, or using a few monolayers of Si grown on GaP as an orientation-reversal layer  \cite{matsushita2007epitaxial,matsushita2009quasi}. An important figure of merit for the fabrication of OP-GaP is domain fidelity, i.e. the ratio between the duty cycle on the surface of the OP-GaP crystal with respect to the duty cycle originally defined in the seed layer. Among the three approaches mentioned above, the highest domain fidelity was demonstrated in OP-GaP fabricated on OP-GaAs seeds\cite{schunemann2021continuous}. Nevertheless, efficient nonlinear conversion processes were only observed in bulk OP-GaP crystals grown by hydrid vapour phase epitaxy (HVPE) for free space quasi-phase matched conversion \cite{rutkauskas2020supercontinuum}. To the best of our knowledge, the only result published on OP-GaP waveguides was based on Si orientation-reversed seeds with limited domain fidelity and dealt with parametric fluorescence \cite{matsushita2009quasi,wei2018performance}.

In the present contribution, an alternative approach to the fabrication of OP-GaP is proposed to produce crystals that are well-suited to the fabrication of shallow-ridge waveguides. Contrary to previous approaches, an OP-GaP seed is obtained by wafer bonding GaP membranes epitaxially grown on GaAs. The seed is then grown to the desired thickness using metal-organic vapor-phase epitaxy. Transmission electron microscopy micrographs reveal that the OP-GaP crystals exhibit a domain fidelity greater than \SI{95}{\percent} from the very first few nanometer of regrowth on the seed and all the way to the surface. These OP-GaP crystals are processed into suspended shallow-ridge waveguides and used to demonstrate continuous-wave second harmonic generation (SHG) {\color{black}with a conversion efficiency} of \SI{200}{\percent\per\watt\per\centi\meter\squared} at telecom wavelengths. These results constitute a significant step forward in the field of OP-GaP, not only in terms of conversion efficiency, but also in the use of a guided geometry instead of free propagation in a bulk crystal and in the demonstration of continuous-wave frequency conversion at power levels relevant to practical applications.

The paper is structured as follows: first, the fabrication process of OP-GaP on GaAs is described and the material properties of the OP-GaP crystals are investigated. Then, the design and fabrication of suspended air-clad OP-GaP waveguides is described. The waveguides are subsequently optically characterized to measure their conversion efficiency and acceptance bandwidth, investigate thermal dissipation in the waveguide as a function of input power and explore the robustness of the waveguides to polarisation diversity. Finally, the results are discussed in the context of performances reported for OP-GaP and other nonlinear platforms.

\section{Experimental\label{sec:exp}}
\subsection{OP-GaP growth\label{sec:opgapfab}}
\begin{figure}[ht!]
    \centering
	\includegraphics[width=0.9\textwidth]{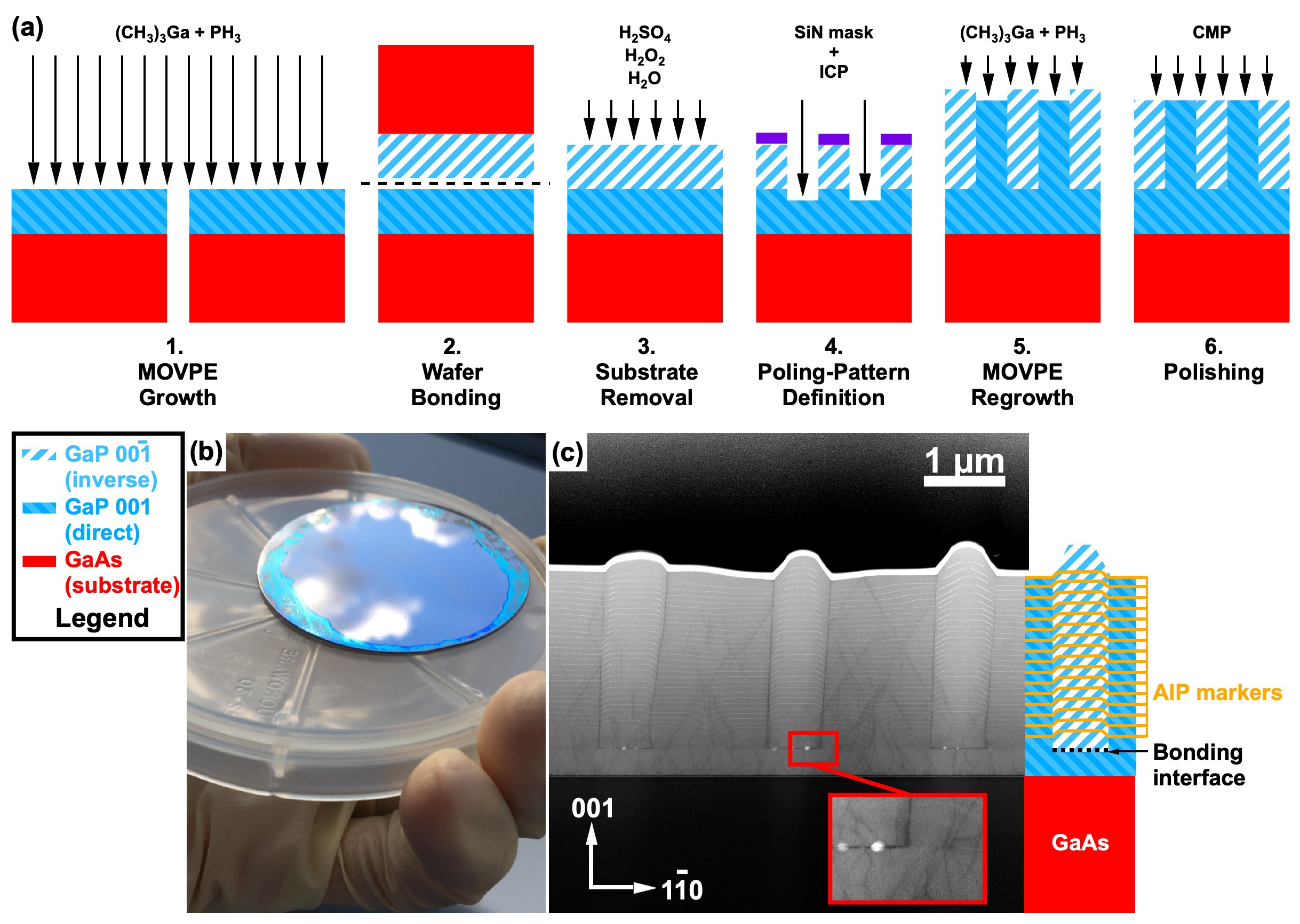}
	\caption{\label{fig:opgap_fab} (a) Fabrication steps of the OP-GaP  seed: MOVPE is used to grow GaP layers on 2-inches GaAs substrates. These are then wafer bonded together and one GaAs substrate is removed. The top layer thus becomes inverted. The photograph in (b) shows  the surface of such a sample after step 3. The surface  is mirror-like reflecting the cloudy sky, indicating excellent surface quality of the transferred GaP. The two GaP orientations exhibit different growth rates, as revealed in the HAADF-STEM micrograph of a control OP-GaP sample shown in (c). {\color{black}The inset here shows the bonding interface of the seed crystal, showing good reconstruction with few pockets of oxide, smaller than \SI{100}{\nano\meter} in diameter.} A last polishing step is used, to level the OP-GaP surface and remove this source of dispersion. Finally, the HAADF-STEM micrograph also shows that OP-GaP obtained in this manner has a domain fidelity of more than \SI{95}{\percent}, with domain boundaries almost perfectly parallel to each other throughout the sample thickness.}
\end{figure}

The fabrication steps of the OP-GaP seed and subsequent growth of  OP-GaP on GaAs crystals are summarized in Fig.~\ref{fig:opgap_fab}~(a). The fabrication involves six steps in total. First, thin GaP layers are epitaxially grown on 001-oriented GaAs substrates using metal-organic vapor-phase epitaxy (MOVPE). Two such GaP membranes on GaAs are fused together using direct wafer bonding. The two membranes are fused along their respective \hkl[001] surface, similarly to the process described for GaAs in References~\cite{Grisard2012,Grisard2012pssc}. The top template having been flipped in the process, it is now \hkl{00-1}-oriented in the referential of the bottom template. In the third step, the top GaAs wafer is selectively etched away in wet chemistry, using a \ce{H2SO4}:\ce{H2O2}:\ce{H2O} solution. This solution is perfectly As/P selective and the GaP layer serves as its own etch-stop layer. The photograph in Fig.~\ref{fig:opgap_fab}~(b), shows the surface of a sample after Step 3. The surface is mirror-like, reflecting the cloudy sky. The inset shows an image from a thermal camera. The central portion is homogeneous, indicating that there are no macroscopic air-voids trapped between the two wafer-fused GaP membranes. In Step 4, a two-level e-beam lithography is used to define the orientation-reversal period using a SiN mask. Chlorine-based inductive coupled plasma (ICP) etching is used to selectively etch away the inverse GaP and reveal the buried direct GaP. The \ce{SiN} mask is then removed. The resulting sample is the OP-GaP seed.

This OP-GaP seed is subsequently grown to the desired OP-GaP thickness using MOVPE (Step 5). This technique is preferred over hydride vapor-phase epitaxy (HVPE) used elsewhere for bulk OP-GaP and OP-GaAs growth, as it offers better control over thickness for crystals that are up to \SI{5}{\micro\meter} thick. A cross-section of a control OP-GaP crystal obtained after Step 5, observed using high-angle annular dark-field scanning transmission electron microscopy (HAADF-STEM), is shown in Fig.~\ref{fig:opgap_fab}~(c). {\color{black}This control sample has a period of \SI{2}{\micro\meter} and a duty cycle of {\color{black}one quarter inverse to three quarters direct GaP}. The period and duty cycle differ from those of sample for SHG, to increase the number of domains observable in TEM.} Three inverse GaP domains, surrounded by direct GaP are visible in the micrograph. The inverse GaP is readily distinguished by the pyramidal features present at the surface of the sample. These features are due to a difference in the growth rate between the inverse and direct GaP. The inverse GaP grows at a rate \SI{10}{\percent} greater than the direct GaP, as revealed by the regularly spaced AlP markers placed in this control sample during Step 5. This difference results in a modulation of the surface thickness, tantamount to a Bragg scatterer. At longer wavelengths, this roughness would not be significant. At telecom wavelengths, however, it can cause significant diffusion losses. A final polishing step is, therefore, implemented (Step 6). After this chemical mechanical polishing (CMP) step, the RMS surface roughness is reduced to \SI{1}{\nano\meter} \cite{Saleem-Urothodi:2020tl}. Finally, the HAADF-STEM cross-section in Fig.~\ref{fig:opgap_fab}~(c) reveals one more important feature of the fabrication process proposed in the present contribution: the domain boundaries between inverse and direct GaP are almost perfectly parallel to the \hkl{001} direction, all throughout the thickness of the OP-GaP crystal, {\color{black}yielding an average domain fidelity of \SI{95}{\percent} over all the domains observed in the TEM lamella (\SI{20}{\micro\meter} in total length)}. 

\subsection{Waveguide Design and Fabrication\label{sec:design}}

\begin{figure}[!ht]
    \centering
	\includegraphics[width=0.7\textwidth]{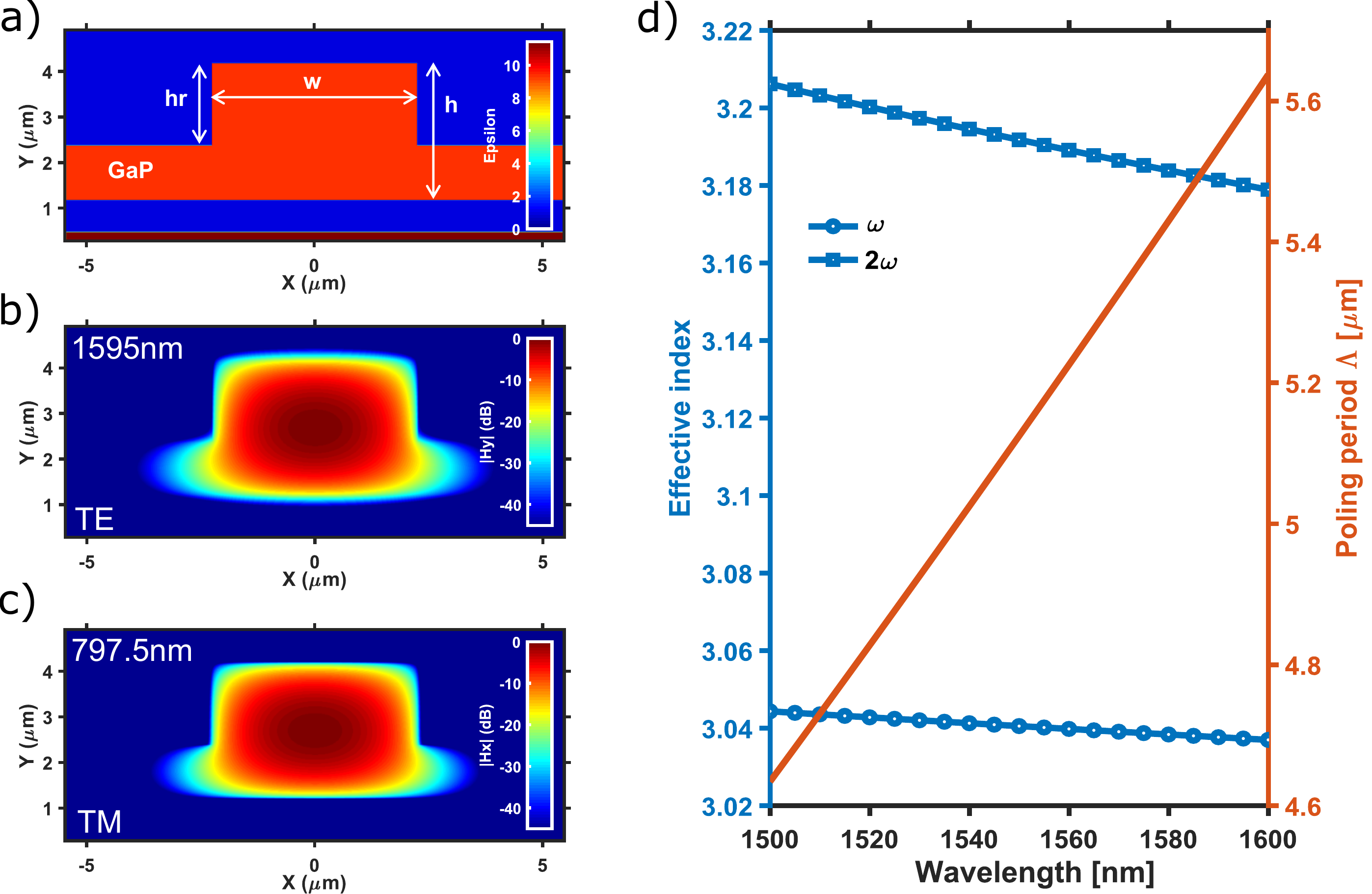}
	\caption{\label{fig:modeprofiles_efficiency}  (a) index profile of the air-clad suspended shallow-ridge waveguide GaP described in the text. (b) Calculated magnetic H-field distribution along y for the fundamental mode in transverse electric (TE) polarisation and (c) magnetic field distribution along x for the generated second harmonic in transverse magnetic (TM) polarisation. The two modes are shown to nearly perfectly overlap. (d) Numerical calculations of the effective indices for the fundamental ($\omega$) and generated second-harmonic ($2\omega$) modes. Also plotted is the period $\Lambda$ for the quasi-phase matching as a function of wavelength (right-hand axis).}
\end{figure}

An air-clad, suspended, shallow-ridge waveguide design was chosen to demonstrate SHG at telecom wavelengths in OP-GaP. This design ensures a maximum overlap between the fundamental and second harmonic modes, while minimizing optical losses. The total height $h$ of the suspended waveguide is \SI{3}{\micro\meter}, the etch-depth $h_{r}$ is \SI{1.6}{\micro\meter}, and the width $w$ of the ridge is \SI{4.5}{\micro\meter}. The index profile of this waveguide is shown in Fig.~\ref{fig:modeprofiles_efficiency}~(a). Given these parameters, the field distributions for the transverse-electric (TE) polarized fundamental mode at \SI{1595}{\nano\meter} and its transverse-magnetic (TM) polarized second harmonic at \SI{797.5}{\nano\meter} were computed using Finite Difference Solver~\cite{fallahkhair2008vector} and are represented in Fig.s~\ref{fig:modeprofiles_efficiency}~(b) and \ref{fig:modeprofiles_efficiency}~(c), respectively. The TE and TM modes are shown to nearly perfectly overlap.

The effective indices for both modes as a function of wavelength were also computed. The plot in Fig.~\ref{fig:modeprofiles_efficiency}~(d) shows the result of this calculation for wavelengths between \SI{1500}{\nano\meter} and \SI{1600}{\nano\meter}. The curve with the dot markers corresponds to the effective index of the fundamental mode, while the curve with the squares corresponds to the second harmonic. Using these effective indices, the dependence of the poling period $\Lambda$, defined for Type-I SHG as $2\pi/\Lambda = k_{2\omega}-2k_{\omega}$, on the wavelength was also calculated. The orange solid curve in Fig.~\ref{fig:modeprofiles_efficiency}~(d) represents the dependence of $\Lambda$ with the wavelength. The plot reveals that the poling period needs to be between \SI{4.6}{\micro\meter} and \SI{5.6}{\micro\meter} for operation in the telecom window. Hence a poling period of \SI{5.5}{\micro\meter} was chosen, i.e. an inversion of the orientation of GaP every \SI{2.75}{\micro\meter}.

Using the above design parameters, OP-GaP on GaAs templates are fabricated and processed to realize suspended shallow-ridge waveguides. The OP-GaP on GaAs templates are obtained according to the process-flow described in Section~\ref{sec:opgapfab}, with only one difference: the bottom-wafer also contains a \SI{700}{\nano\meter} thick \ce{Al_{0.85}Ga_{0.15}As} etch-stop layer underneath the GaP membrane. The total OP-GaP height is \SI{3}{\micro\meter}.

\begin{figure}[!ht]
     \centering
 	\includegraphics[width=0.9\textwidth]{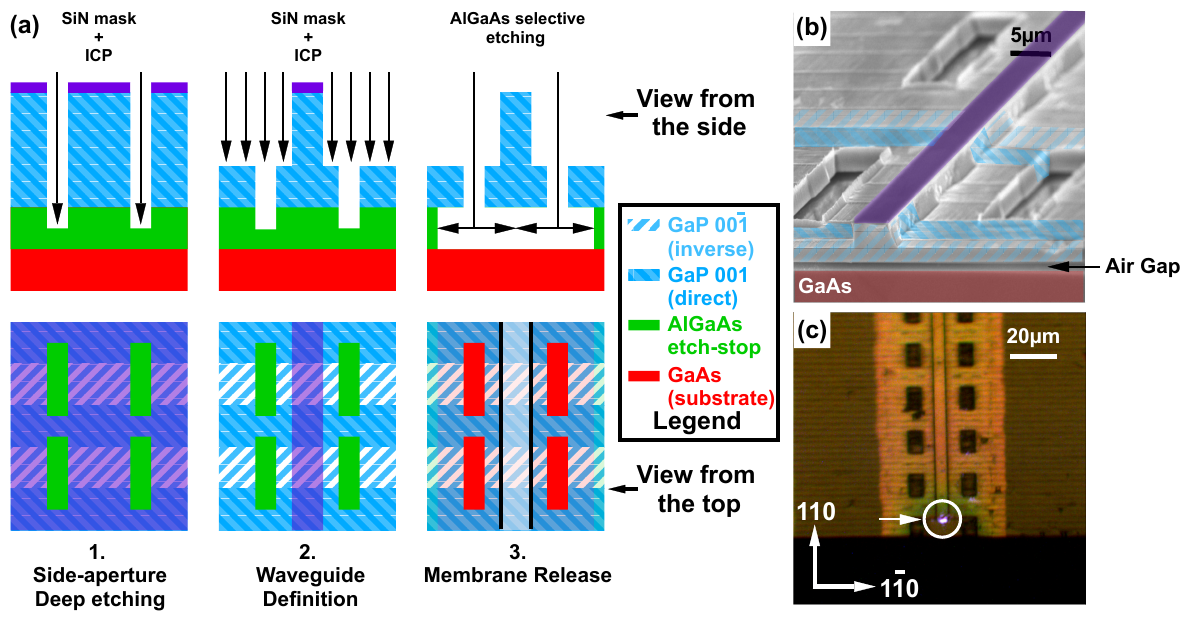}
 	\caption{\label{fig:waveguide_processing}  (a) Processing steps for the fabrication of the orientation-patterned GaP suspended waveguide. Step 1: Deep trenches are etched to reveal the buried AlGaAs. Step 2: A \SI{1.6}{\micro\meter} deep by \SI{4.5}{\micro\meter} wide shallow-ridge waveguide is etched into the orientation-patterned GaP. Step 3: the buried AlGaAs is etched away using wet-etching, suspending the waveguide. (b) {\color{black}a bird's-eye view of a suspended OP-GaP waveguide at the center of the \SI{50}{\milli\meter} wafer, acquired in a SEM; a periodic modulation of the thickness at the waveguide's sidewalls is visible}. This modulation is due to the difference in the undercut during etching of the direct and inverse GaP. (c) Visible-light top view of the waveguide (right), acquired in situ during optical measurements. The bright spot on the bottom edge of waveguide (inside the white circle) is light from the the second harmonic at a wavelength of \SI{798.2}{\nano\meter} or, equivalently \SI{375.6}{\tera\hertz}.}
\end{figure}

The processing steps of the suspended OP-GaP waveguide are schematically depicted in Fig.~\ref{fig:waveguide_processing}~a. The first step here is to etch \SI{5}{\micro\meter} by \SI{10}{\micro\meter} apertures on the sides of the waveguide, all the way into the buried \ce{Al_{0.85}Ga_{0.15}As} layer. These apertures serve as vias to suspend the waveguide in the final processing step. Then, the \SI{1.6}{\micro\meter} deep by \SI{4.5}{\micro\meter} wide waveguide is etched into the OP-GaP layer. Finally, the waveguide is suspended by selectively removing the \ce{Al_{0.85}Ga_{0.15}As} using wet etching. After processing, the sample is cleaved, creating \SI{1}{\milli\meter}-long waveguides.

Fig.~\ref{fig:waveguide_processing}~b is
a scanning electron microscope (SEM) micrograph, showing a bird's-eye view of the suspended waveguide. Lines running perpendicularly to the waveguide reveal the periodic alternation of the direct and inverse GaP domains. The sidewalls of the waveguide show a periodic modulation in width, that {\color{black} is related to the masking step in the processing of the waveguide}. Finally, Fig.~\ref{fig:waveguide_processing}~c is a top view of the guide, recorded with the visible-light camera of the optical setup used in subsequent sections. The white spot at the end of the waveguide in this video capture is due to diffusion of the generated second harmonic - $f=$~\SI{375.6}{\tera\hertz} or $\lambda=$~\SI{798.2}{\nano\meter} - at maximum conversion efficiency.

\section{Results\label{sec:experiment}}

Experiments to measure the efficiency of SHG in the suspended shallow ridge OP-GaP waveguides described above were carried out using the following experimental setup: a tunable continuous-wave laser from Santec (model TSL510) is used as a source. The laser light is amplified using an L-band Erbium-doped amplifier from Lumibird (CEFA-L-HG-SM-40). After amplification, an electronic variable optical attenuator from OzOptics (DD-100) is used to control the input power injected into the waveguide. A 1/99 coupler followed by a powermeter (Thorlabs) is used to monitor the input power. The attenuator is followed by a polarisation controller for the injected beam. Light is injected and collected from the waveguide through microscope objectives (magnification x50, numerical aperture 0.65). A free-space polarizer at the waveguide exit is used to collect light in either TE or TM polarisation. The polarizer is set to TE to measure the transmitted fundamental,  and to TM to measure the generated second harmonic. The focus of the collection objective is also adjusted between the two types of measurements, to compensate for achromaticity. Collected light is analyzed using a Yokogawa (AQ6370D) optical spectrum analyzer (OSA). The group index and losses have been analyzed using Fabry-Pérot analysis (see Supporting Information for additional information).

\begin{figure}[!ht]
\centering
\includegraphics[width=0.9\textwidth]{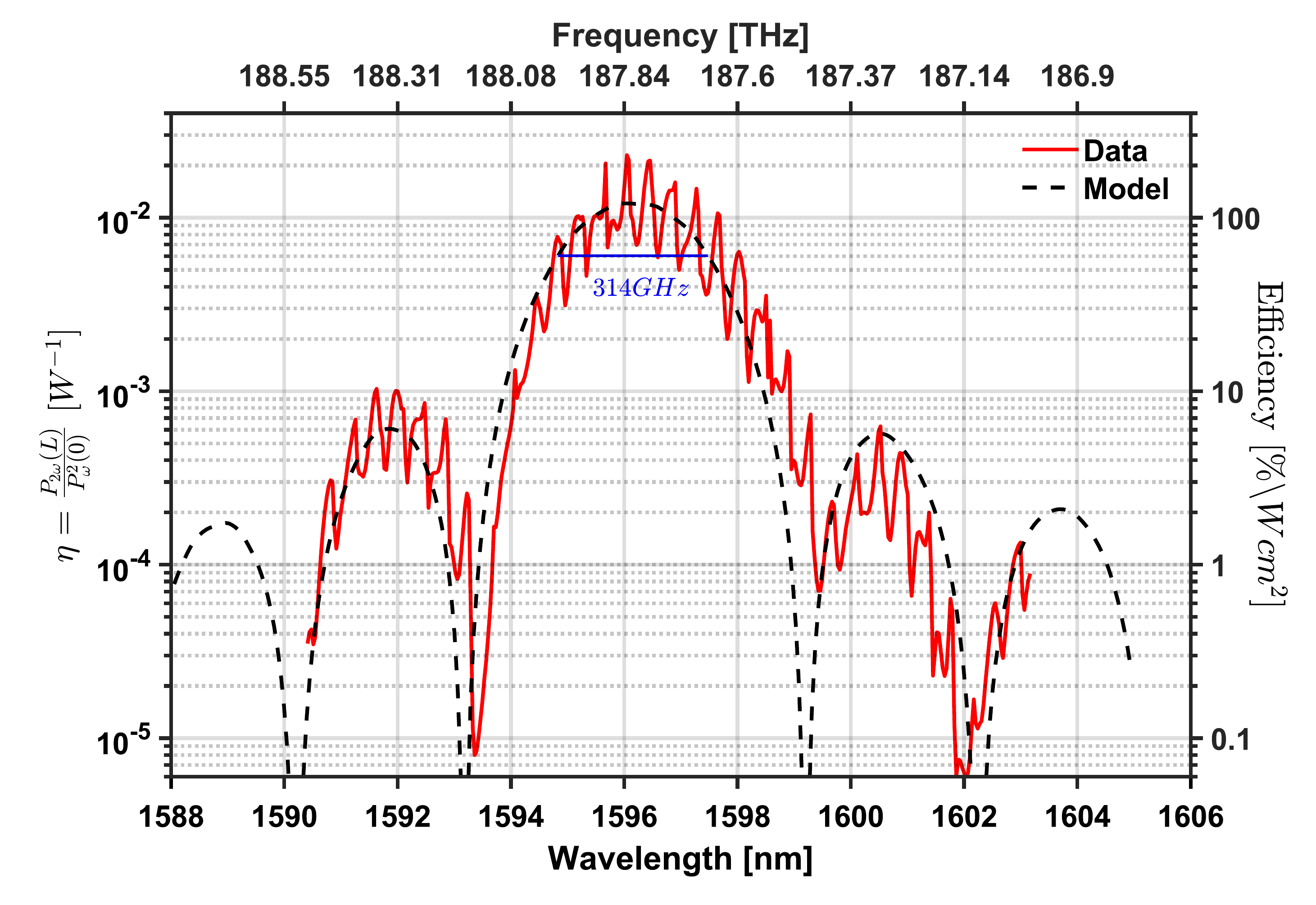}
\caption{\label{fig:SHG_spectral_aceptance}  Conversion efficiency for Type-I SHG as a function of the pump wavelength (bottom) and frequency (top) of the fundamental. The red solid curve represents the experimental data. The envelope of the curve follows a squared cardinal sine function, in good agreement with the model from Equation~\ref{eq:1} (dashed black curve). The peak conversion efficiency is \SI{200}{\percent\per\watt\per\centi\meter\squared} and the \SI{3}{\decibel} bandwidth is \SI{314}{\giga\hertz} or, equivalently,  \SI{2.67}{\nano\meter}. Both the conversion efficiency and the bandwidth are the highest reported for OP-GaP waveguides.}
\end{figure}

The measured internal second-harmonic conversion efficiency $\eta$ in the OP-GaP waveguides, defined as $\eta = P_{2\omega}(L)/P^2_{\omega}(0)$, is represented in Fig. \ref{fig:SHG_spectral_aceptance} for Type-I SHG. The experimental curve is shown in the solid red line and corresponds to an incident pump power of \SI{70}{\milli\watt}. $P_{2\omega}(L)$ represents the power of the second harmonic collected at the output of the waveguide of length $L$, and $P_{\omega}(0)$ the power of the pump at the entrance of the waveguide. The measured power is corrected for input and output coupling losses, estimated at \SI{3.75}{\decibel} each (including Fresnel and microscope objective). The lineshape of the plot follows a $sin(x)/x$ function, characteristic of the spectral acceptance of the SHG process. It is modulated by periodic fluctuations that are due to Fabry-Pérot interference. The envelope of the conversion efficiency in orientation-patterned channel waveguides follows a squared cardinal sine function and can be modelled following Reference \cite{pliska1998linear}. This model takes into account propagation losses and expresses $\eta$ as follows:

\begin{equation}
\centering
\eta = \frac{2\omega^2}{\epsilon_0c^3} \frac{d^2_\textnormal{eff}}{n_\omega^2n_{2\omega}}L^2\mathcal{H}\Gamma.
\label{eq:1}
\end{equation}

Here, $\omega$ and $2\omega$ are the angular frequencies of the fundamental and the second harmonic modes, respectively, $n_\omega$ and $n_{2\omega}$ their respective effective refractive indices, $\epsilon_0$ the vacuum permittivity, and $c$ the speed of light. The effective non-linear coefficient is $d_\textnormal{eff}=\frac{2}{\pi}d_{36}$, following the normalization associated to quasi-phase matching and considering the symmetries of the GaP crystal \cite{kleinman1962nonlinear}, with a $d_{36}$ of \SI{50}{\pico\meter\per\volt} in GaP at \SI{1.55}{\micro\meter}.

The mismatch function $\mathcal{H}$ is:

\begin{equation}
\label{eq:2}
\centering
\mathcal{H} = \frac{\Big(\sinh\big(\Delta\alpha L/2\big)\cos\big(\Delta\beta L/2\big)\Big)^2+\Big(\cosh\big(\Delta\alpha L/2\big)\sin\big(\Delta\beta L/2\big)\Big)^2}{\Big(\big(\Delta\alpha^2+\Delta\beta^2\big)L^2/4\Big)}\times e^{-\Delta\alpha L},
\end{equation}
with $\Delta\alpha$ and $\Delta\beta$ the loss and phase mismatch respectively. The loss mismatch is defined as $\Delta\alpha=\alpha_{2\omega}/2-\alpha_\omega$, with $\alpha_{2\omega}$ and $\alpha_{\omega}$ the respective losses for the fundamental and the second harmonic. The phase mismatch is defined as $\Delta\beta=\beta_{2\omega}-2\beta_{\omega}-2\pi/\Lambda$ with the wavevectors $\beta_{\omega} =\omega n^{}_
{\omega}/c$ and $\beta_{2\omega} =2\omega n^{}_{2\omega}/c$. Here, $\Lambda$ is the poling period, as defined in Section~\ref{sec:design}, and $L$ is the length of the waveguide, which is \SI{1}{\milli\meter} in the present experiments.
Finally, the overlap integral $\Gamma$ is:
\begin{equation}
    \label{eq:gamma}
    \centering
\Gamma = \frac{\Big(\iint_{\mathbf{R}^2} E^2_\omega(x,y)E_{2\omega} \,\textnormal{d}x\,\textnormal{d}y\Big)^2}{\Big(\iint_{\mathbf{R}^2}E^2_\omega(x,y) \textnormal{d}x\,\textnormal{d}y\Big)^2\iint_{\mathbf{R}^2}E_{2\omega}^{2}\,\textnormal{d}x\,\textnormal{d}y}.
\end{equation}
Here, $E_{\omega}$ is the electrical field for the fundamental TE mode (along $x$) and $E_{2\omega}$ the electrical field for the second harmonic  TM mode (along $y$).

{\color{black}A very good agreement between the measured and predicted spectral dependence of the conversion efficiency can be observed in Fig.~\ref{fig:SHG_spectral_aceptance}. This agreement was obtained using parameters that were defined above and measured on the sample, except for losses at $2\omega$ and the total thickness of the membrane. In terms of the losses at $2\omega$, a value of $\alpha_{2\omega}$ of \SI{13.6}{\per\centi\meter} was used to match the peak conversion efficiency of \SI{200}{\percent\per\watt\per\centi\meter\squared} (for a more detailed discussion on losses, see Supporting Information). The total thickness of the membrane for the guide was also adjusted, from \SI{3}{\micro\meter} to \SI{2.5}{\micro\meter}, to account for the maximum conversion efficiency occurring at \SI{1596}{\nano\meter}, instead of the targeted \SI{1580}{\nano\meter}. This adjustment accounts for a gradient the thickness of the OP-GaP layer across the \SI{50}{\milli\meter} wafer, induced by the CMP polishing in Step 6 of the fabrication of the OP-GaP seed. The suspended membrane in Fig.~\ref{fig:waveguide_processing}~(b), where the thickness was measured is representative of the center of the wafer, whereas the guide that yielded the highest efficiency is situated at approximately \SI{15}{\milli\meter} from the center.} The  \SI{3}{\decibel} bandwidth is \SI{314}{\giga\hertz} or, equivalently, \SI{2.67}{\nano\meter}.

\begin{figure}[!ht]
    \centering
	\includegraphics[width=1\textwidth]{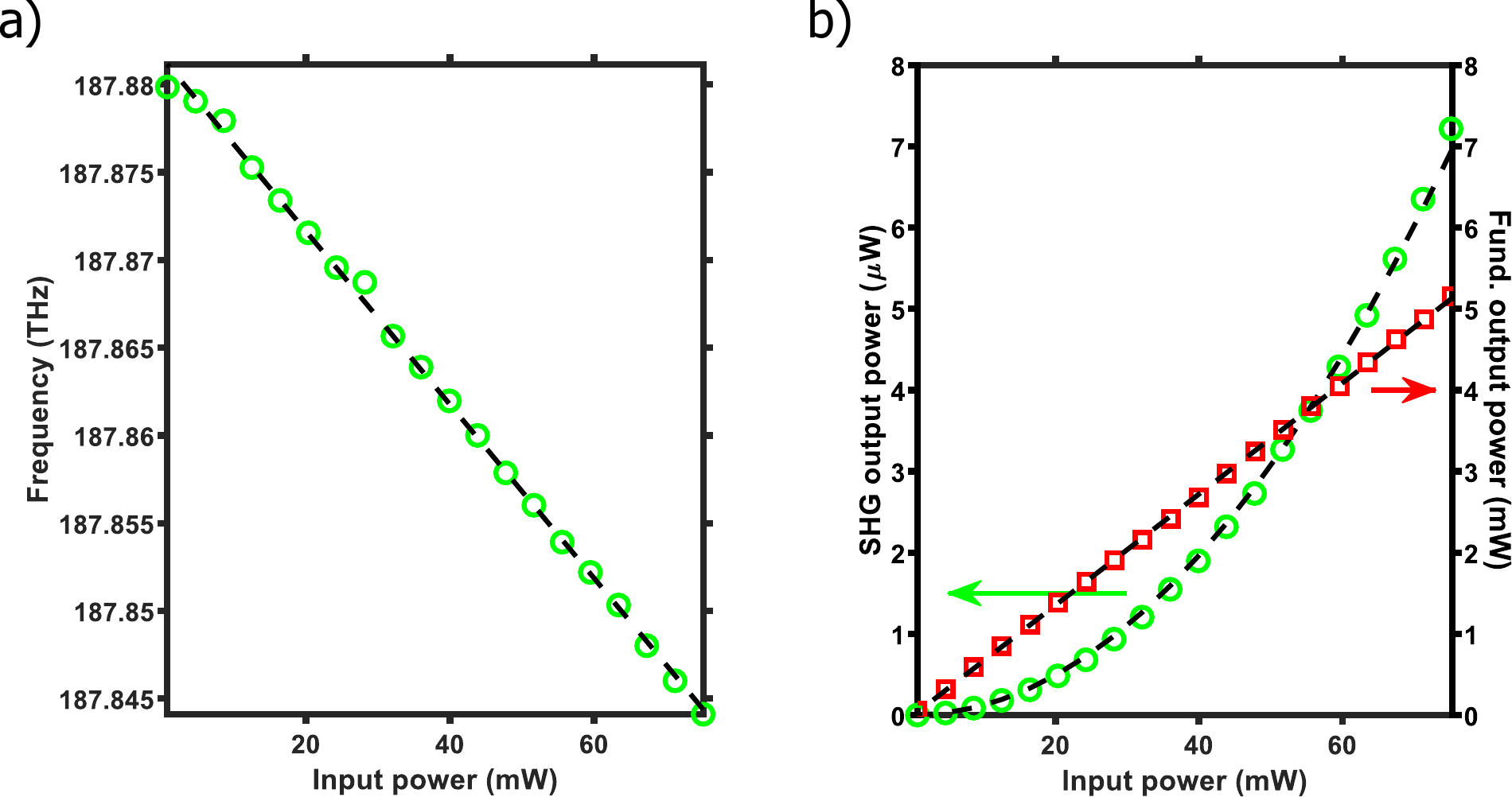}
	\caption{\label{fig:power_dependance}  (a) Plot of frequency at which maximum SHG is observed versus the input power. A linear drift is observed. (b): Plot of the transmitted power for the fundamental (red squares, right-hand axis) and the second harmonic (green circles, left-hand axis) as a function of input pump power. Fits to the two curves are given in dashed black lines. The transmitted power scales linearly for the fundamental and quadratically for the second harmonic, in keeping with theory.}
\end{figure}

Fig. \ref{fig:power_dependance}~(a) demonstrates a linear dependence of the frequency at maximum conversion on the input pump power, with a slope of $\frac{\textnormal{d}f}{\textnormal{d}P}=$\SI{1.14}{\tera\hertz\per\watt} (a detailed derivation is presented in Supporting Information). This is related to an increase of the temperature in the waveguide, due to linear absorption, either at the surface (domain interface or side-walls) or in the bulk (residual doping). Note that the sample is not temperature-stabilized by a Peltier module. From the temperature elevation and after calculating the thermal resistance using finite-element simulations, the absorption coefficient $\alpha$ is \SI{0.27}{\per\centi\meter}, an order of magnitude lower than scattering losses discussed here above.
The plots in Fig.~\ref{fig:power_dependance}~(b) show the transmitted power of the fundamental (red squares, right-hand side axis) and the second harmonic (green circles, left-hand side axis) as a function of the input pump power. The reported values correspond to the maximum conversion efficiency at each value of the input power. The output power scales linearly for the fundamental and quadratically for the second harmonic behaviour, in keeping with theory. Overall, the SHG process is well described by simple models, an indication that the detrimental impact of inhomogeneities is negligible here, and that with exception of propagation losses, the device operates close to its ideal regime.

\begin{figure}[!ht]
    \centering
	\includegraphics[width=0.8\textwidth]{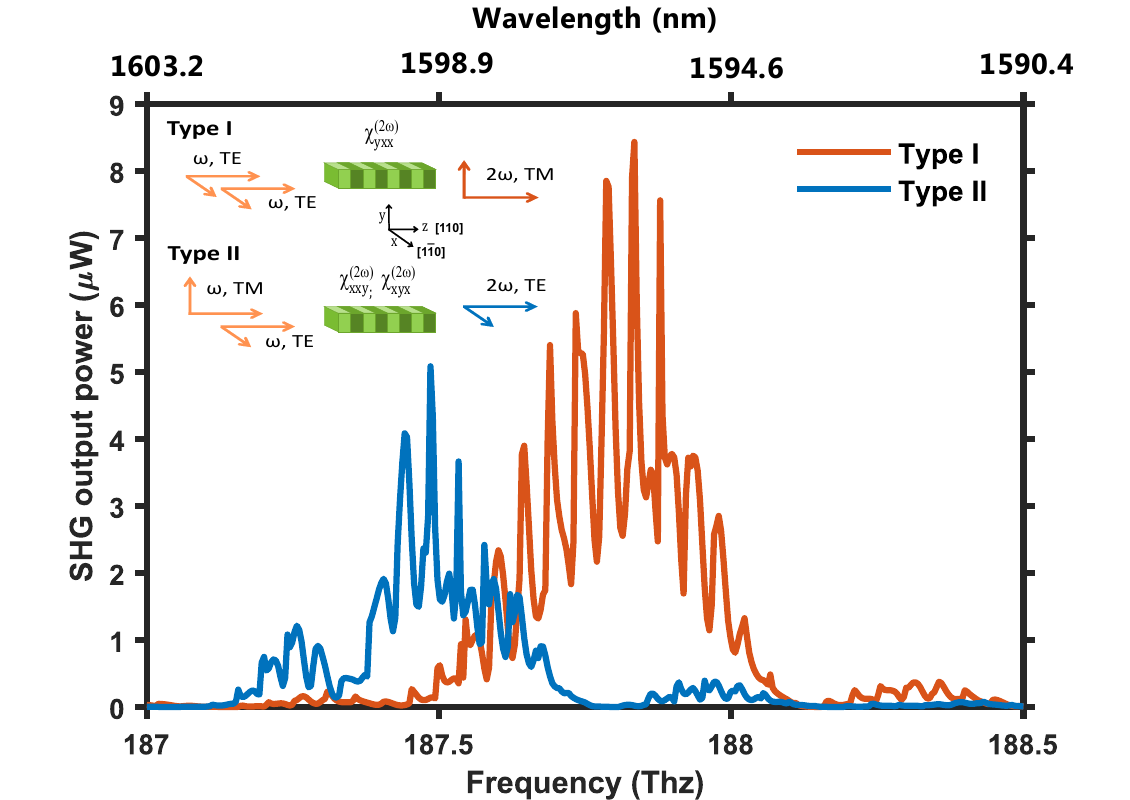}
	\caption{\label{fig:type1typ2} Measured second-harmonic output power (linear scale) versus incident pump wavelength for Type-I and Type-II configurations. The inset shows a schematic representation of the input and output polarisations of fundamental and converted photons for Type-I and Type-II SHG. The output powers in both Type-I and Type-II configurations are comparable{\color{black}, with the Type-II slightly lower due to increased losses for TM ploarisation (see Supporting Information)}. The spectra overlap for frequencies between \SI{187.5}{\tera\hertz} and \SI{188.75}{\tera\hertz}.}
\end{figure}

The OP-GaP waveguides presented here also handle polarisation diversity \cite{oron2010efficient}. Type-II configuration (see schematic on Fig.~\ref{fig:type1typ2}), combining two input photons from orthogonal polarisations (TE and TM), leads to SHG in the TE polarisation. Fig.~\ref{fig:type1typ2} compares the SHG spectra for both Type-I and Type-II configurations for \SI{70}{\milli\watt} incident optical power. The plots in the figure show that the SHG efficiency for both Type I and Type II is comparable{\color{black}, with peak output power for the Type-II configuration about two thirds of the peak output power of Type I. This discrepancy is due to increased losses for the TM polarisation at \SI{1596}{\nano\meter}  (see Supporting Information).} Moreover, their spectral acceptance overlaps partially, making it possible, in principle, to operate both simultaneously. The amount of overlap in the lineshapes is directly related to the slant angle of the sidewalls - here unintentional (see Supporting Information). However, it is possible to adjust the waveguide fabrication process to control this angle and tune the waveguide for  polarisation diversity.

\section{Discussion}

The conversion efficiency reported here, \SI{200}{\percent\per\watt\per\centi\meter\squared}, is the highest to date for OP-GaP, several orders of magnitude higher than other reports in the literature for the same material and second-harmonic generation. This improvement is both due to the quality of the OP-GaP crystal, i.e. the very high domain fidelity, and confinement in the waveguide. In guided geometries, this efficiency is a five-hundred fold improvement over the most recent report of second-harmonic generation in GaP waveguides \cite{Anthur:21}, were modal matching was used. Furthermore, compared to that approach, the second harmonic is generated in the lowest-order TE or TM mode, making fiber-coupling straightforward.

Nevertheless, the efficiency reported here is still significantly lower than that reported for waveguides fabricated in competing nonlinear materials, in particular PPLN\cite{wang2018ultrahigh} and OP-GaAs. A major discriminating factor here are the losses in the waveguide, \SI{7}{\per\centi\meter}. These are an order of magnitude higher than those reported for competing technologies. A detailed analysis of the power-induced spectral drift, reveals that absorption in OP-GaP accounts for only \SI{0.27}{\per\centi\meter} (for a detailed analysis see Supporting Information. The remainder is, therefore, due to scattering in the waveguide, related to the relatively high roughness of the waveguide's sidewalls. Improvements to the fabrication process of the waveguides, in particular etching, will help significantly reduce these losses and a ten-fold improvement can be reasonably expected. Indeed, similar improvements have been obtained in the past for technologically more mature III-V semiconductors, such as GaAs or InP.

A few other factors can help boost the conversion efficiency. The simplest is to increase the total waveguide length. This approach works up to an optimal length $L_\textnormal{opt}$, past which losses overtake the conversion. This optimal length derives from Equation~\ref{eq:1} and follows:
\begin{equation}
    L_\textnormal{opt}=\frac{2}{\alpha_{2\omega}-2\alpha_\omega}\ln\frac{\alpha_{2\omega}}{2\alpha_\omega}.
    \label{eq:optlen}
\end{equation}
The full derivation can be found in Supporting Information. Given the losses measured here, this optimal length is rather \SI{2}{\milli\meter}. Increasing the length of the current waveguides to this length would yield a modest improvement of only \num{1.4}. However, decreasing the losses by a factor two, would already allow the fabrication of waveguides that would be optimal at \SI{4}{\milli\meter}, a common length for such guides in the literature, and yield a conversion efficiency of \SI{1100}{\percent\per\watt\per\centi\meter\squared}. As a reference, if propagation loss were comparable to what reported on more mature semiconductor photonic waveguides, e.g. \SI{0.7}{\per\centi\meter}, the conversion efficiency would rise to \SI{6000}{\percent\per\watt\per\centi\meter\squared} for a length of \SI{4}{\milli\meter}, more than twice that of the current state of the art for PPLN waveguides of the same length.\\

The last option to improve the conversion efficiency is to increase confinement. Calculations show that homothetically reducing the width and height of the waveguide to $0.5\times$\SI{0.6}{\micro\meter\squared} will increase the efficiency by more than an order of magnitude. Furthermore, if ultimate efficiency were the goal one could opt for a resonant device, such as ring resonators \cite{Lu:19}.

In terms of bandwidth, the measured \SI{314}{\giga\hertz} correspond to \SI{2.67}{\nano\meter}, a value that is similar to the largest bandwidths reported in the literature for the state-of-the-art in competing materials. In terms of the targeted application - conversion from the telecom C-band to visible wavelengths - it may be advantageous to further increase this bandwidth. The full C-band is \SI{35}{\nano\meter} wide; conversion of the full band using a single OP-GaP guide, or any competing technology, would require a fifteen-fold increase of the bandwidth. One approach towards broadening the conversion bandwidth consists in shortening the waveguide. Though feasible, this approach comes at the cost of decreased efficiency. It could, therefore, only be envisioned if losses in the OP-GaP waveguides are decreased.

An alternative approach would consist in using chirped waveguides, i.e. waveguides where the QPM period $\Lambda$ is slowly varied over the length of waveguide, thus inducing engineered chirp that broadens the bandwidth \cite{jankowski2020ultrabroadband}. This approach would require a longer waveguide, typically $\SI{10}{\milli\meter}$, and would again be realistic when losses in the OP-GaP waveguides are reduced by a factor five.

Finally, the OP-GaP waveguides presented here exhibit significant Fabry-Pérot interference. This can be detrimental to device performance, especially in the case of full C-Band conversion. This issue can be addressed using anti-reflection  coatings or tapers. Either option is straightforward to integrate in the current design.

\section{Conclusion}

A new approach that produces OP-GaP with very high domain fidelity throughout the crystal has been introduced. These OP-GaP crystals have been used to fabricate suspended shallow-ridge OP-GaP waveguides that allow for continuous-wave second-harmonic generation at telecom frequencies. A conversion efficiency of \SI{200}{\percent\per\watt\per\centi\meter\squared} and an acceptance bandwidth of \SI{2.67}{\nano\meter} are reported. Furthermore, these waveguides exhibit very little thermal drift, even at coupled-power levels as high as several tens of milliwatt. Finally, the OP-GaP waveguides offer the possibility to manage polarisation diversity, as evidenced by the overlapping efficiency curves for Type-I and Type-II configurations.
There is still a significant margin for improvement in the performance of these OP-GaP waveguides, both in terms of conversion efficiency and acceptance bandwidth. This improvement can be obtained by increasing the confinement or optimizing the waveguide fabrication process. More importantly, these suspended OP-GaP waveguides can be readily transferred onto Si photonic integrated circuits using pick-and-place technologies such as micro-transfer printing \cite{billet2020gallium}. They thus constitute a new bridge over the gap between mature and cost-effective visible detectors and near-infrared PICs and optical processing needs at telecom wavelengths and above, including spontaneous parametric down conversion (SPDC) for emission of entangled photon pairs based on efficient quasi-phase matched designs.

\begin{acknowledgement}

The authors would like to acknowledge help from A. Marceau with additional numerical simulations.

\end{acknowledgement}

\section{Funding Sources}

The present work has been supported by the French Renatech network and by the National Research Agency (ANR-17-CE24-0019). F. Talenti received funding from the European Union’s Horizon 2020 Research and Innovation Program under the Marie Skłodowska-Curie grant agreement No. 814147.

\begin{suppinfo}
See Supporting Information for additional results on losses, thermal drift, optimal interaction length, and further discussion on Type-I and Type-II conversion efficiencies and how they are affected by slanted waveguides.
\end{suppinfo}

\bibliography{./SHG_References}

\newpage

\section{For Table of Contents Only}

Scanning electron microscopy micrograph showing a bird's-eye view of Orientation-Patterned Gallium Phosphide processed into a suspended air-clad waveguide for second-harmonic generation experiments.

\begin{figure}
    \centering
    \includegraphics[]{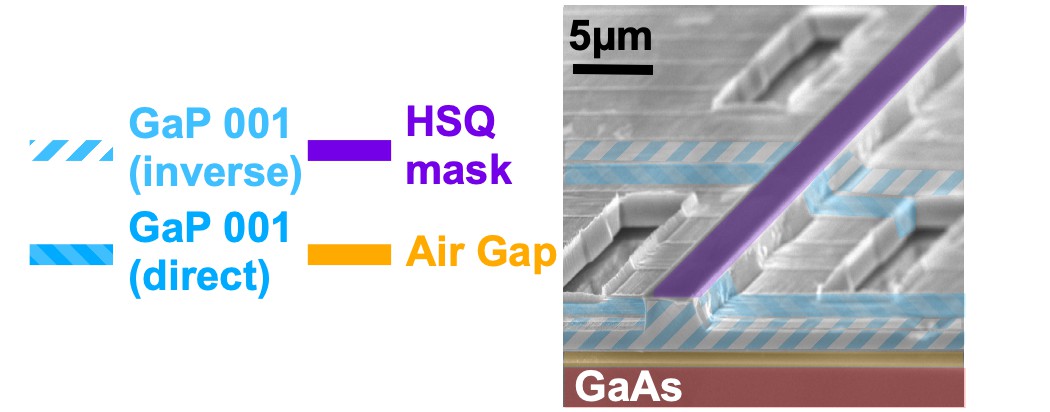}
\end{figure}

\end{document}